\documentclass[letterpaper, 10 pt, conference]{ieeeconf}  

\IEEEoverridecommandlockouts                              

\usepackage{graphics} 
\usepackage{graphicx}
\usepackage{epsfig} 
\usepackage{times} 
\usepackage{amsmath} 
\usepackage{amssymb}  
\usepackage{mathtools}
\usepackage{nccmath}
\usepackage{color,url}
\usepackage{neuralnetwork}
\usepackage{xpatch}
\makeatletter
\xpatchcmd{\linklayers}{\nn@lastnode}{\lastnode}{}{}
\xpatchcmd{\linklayers}{\nn@thisnode}{\thisnode}{}{}

\usepackage {tikz}
\usetikzlibrary {positioning}
\definecolor {processblue}{cmyk}{0.96,0,0,0}
\usetikzlibrary{shapes,arrows}
\usetikzlibrary{calc}
\usetikzlibrary{matrix,chains,decorations.pathreplacing}

\usepackage{amsthm}
\usepackage[utf8]{inputenc}
\usepackage[T1]{fontenc}

\usepackage[ruled,linesnumbered]{algorithm2e}
\usepackage{algpseudocode}

\newcommand{\trp}{\intercal}

\DeclareMathOperator*{\minimize}{min\,}

\input{mysymbol.sty}
\usepackage{needspace}

\theoremstyle{definition}

\theoremstyle{plain}

\theoremstyle{remark}

\makeatletter
\let\NAT@parse\undefined
\makeatother
\usepackage[hidelinks]{hyperref}
\usepackage{xcolor}
\hypersetup{
    colorlinks,
    linkcolor={red!50!black},
    citecolor={blue!50!black},
    urlcolor={blue!80!black}
}

\title{\LARGE \bf Optimal Resource Allocation in Wireless Control Systems \\ via Deep Policy Gradient
}

\author{Vinicius Lima$^{1}$, Mark Eisen$^{2}$, Konstantinos Gatsis$^{1}$ and Alejandro Ribeiro$^{1}$
\thanks{Supported by Intel Science and Technology Center for Wireless Autonomous Systems and ARL DCIST CRA W9111NF-17-2-0181. $^{1}$Electrical and Systems Engineering, University of Pennsylvania, Philadelphia, PA: 
        {\tt\small \{ \href{mailto:vlima@seas.upenn.edu}{vlima}, \href{mailto:kgatsis@seas.upenn.edu}{kgatsis}, \href{mailto:aribeiro@seas.upenn.edu}{aribeiro} \}@seas.upenn.edu}. 
        $^{2}$ Intel Corporation, Hillsboro, OR: {\tt\small \href{mailto:mark.eisen@intel.com}{mark.eisen@intel.com} }. } %
}

\begin{document}
\maketitle
\thispagestyle{empty}
\pagestyle{empty}

\begin{abstract}
In wireless control systems, remote control of plants is achieved through closing of the control loop over a wireless channel. As wireless communication is noisy and subject to packet dropouts, proper allocation of limited resources, e.g. transmission power, across plants is critical for maintaining reliable operation. In this paper, we formulate the design of an optimal resource allocation policy that uses current plant states and wireless channel states to assign resources used to send control actuation information back to plants. While this problem is challenging due to its infinite dimensionality and need for explicit system model and state knowledge, we propose the use of deep reinforcement learning techniques to find neural network-based resource allocation policies. In particular, we use model-free policy gradient methods to directly learn continuous power allocation policies without knowledge of plant dynamics or communication models. Numerical simulations demonstrate the strong performance of learned policies relative to baseline resource allocation methods in settings where state information is available both with and without noise.
\end{abstract}

\section{INTRODUCTION}
Wireless communication networks are frequently used to exchange data between plants, sensors and actuators in control systems. The use of wireless networks in lieu of wired communication makes the installation of components easier and maintenance more flexible, but also adds particular challenges to the design of control and communication policies  \cite{hespanha_survey_2007, park_wireless_2018}.
Wireless networks are in general more noisy than their wired counterparts \cite{schenato_foundations_2007}, whereas reliable operation of control systems over a wireless medium demands rapid communication and low message error rates --- requirements in turn constrained by the limited resources available in that network. 
It is natural in this setting to look for an optimal way to distribute the resources available in the network among the plants sharing that communication medium. 
Finding an optimal solution to this resource allocation problem, however, is often computationally hard, and allocation in WCSs is usually designed via heuristics and ad-hoc methods \cite{Lamnabhi-Lagarrigue_2017}.  
To overcome the hardness of the problem, we leverage reinforcement learning techniques to find a data-driven resource allocation policy. 

Reinforcement learning gives a mathematical representation to the idea that learning occurs in interaction with the environment: an agent performs an action, receives a reward from the environment, and keeps exploring its surroundings while optimizing some cumulative performance metric. This fairly straightforward structure makes the framework amenable to many engineering problems, particularly those in which explicit model information is unavailable. 
Common reinforcement learning methods fall into two categories: value-based methods, which rely on the calculation of the value function and then compute the corresponding optimal policy, and policy-based methods, which search directly for a policy instead \cite{sutton_reinforcement_learning,busoniu_reinforcement_2018,buoniu_comprehensive_2008}. Although potentially more sensitive to noise, policy gradient methods allow us to model continuous allocation functions; hence our  focus on this class of algorithms for continuous resource allocation problems. In particular, in this work we develop the use of resource-constrained policy gradient methods for performing resource allocation in a network of wireless control systems.

Resource allocation in wireless networks revolves around issues such as power consumption, scheduling, low latency, and high reliability, while taking into account stochastic noise and rapid variability in the communication channel known as fading  
\cite{fattah_leung_scheduling_wl_2002, Ribeiro2012}. 
Those problems involve optimizing a performance metric over a function, resulting in an infinite dimensional problem that is usually hard to solve. 
That formulation, however, resembles a statistical learning problem \cite{Ribeiro2012}, which allows one to treat resource allocation in a model-free data-driven fashion \cite{eisen_learning_2018, liang_opc_dnns}. 

When  control plants  share a  communication network, we also need to take into account the dynamic behavior of each plant as well as stability issues. 
Stability analysis and design of controllers for networked control systems are  considered in \cite{walsh_stability_2002, nesic_input-output_2004, tabbara_input-output_2005}, among others. A classical review on the topic is  \cite{hespanha_survey_2007}; for a recent overview of issues  and algorithms in network design of wireless control systems in particular, we refer the reader to \cite{park_wireless_2018}. 
In this case we need to consider the use of bandwidth and power resources in the wireless network. That is  the problem works on resource allocation and scheduling tackle, cf. e.g. \cite{rehbinder_scheduling_2004, mo_sensor_2011, gatsis_opportunistic_2015, charalambous_resource_2017, eisen_control_2018}. 
This results in a hard optimization problem in which resources are allocated depending upon both the state of the control plants and state of the wireless channel; finding exact solutions invariably requires accurate model and state information. Recent advances in machine learning have motivated the use of data-driven approaches ---i.e. reinforcement learning--- for  scheduling  \cite{demirel_deepcas:_2018,leong_deep_2018, baumann_drl_etc}. In \cite{demirel_deepcas:_2018,leong_deep_2018}, authors utilize Deep Q-Networks (DQN) parameterization to learn a scheduling algorithm based on value iteration. Value-based methods are adequate for discrete scheduling actions, but unsuitable for learning the continuous action spaces of resource allocation problems we consider in this paper. 
In \cite{baumann_drl_etc} the authors explore actor-critic algorithms to learn communication and control policies in event-triggered wireless control systems. Such existing works, however, only consider discrete scheduling actions and use simple communication models that do not take wireless fading into account. 

In this paper we propose the use of policy gradient methods to find optimal resource allocation policies for wireless control systems under power constraints. We cast the resource allocation problem in WCSs as a constrained reinforcement learning problem. This can be solved in continuous action spaces via policy gradient methods that respect wireless resource constraints. We further propose the use of neural networks to parameterize a policy that uses current plant and wireless fading state information to allocate wireless resources. 
 Numerical results demonstrate the strong performance of such resource allocation policies over heuristic benchmarks in settings with both perfect and noisy state information available. Throughout the paper uppercase letters refer to matrices and lowercase letters to vectors. Positive (semi)definiteness of a matrix is indicated by $X (\geq) > 0$. $\mathbb{R}$ and $\mathbb{N}$ stand for the set of real and natural numbers. 

\section{Resource Allocation in  Control Systems}
\label{sec:prob_formulation}
Consider a system made up by $m$ independent plants sharing a common wireless medium as in  Figure \ref{fig:wl_control_sys}. At each time instant a plant samples its state  and send this information to a common access point (AP) containing a centralized controller.  
We assume the dynamics of each plant $i$ can be approximately represented by a linear model affected by some random noise  standing for eventual disturbances and linearization errors or unmodeled dynamics, such that 
\begin{equation}
 x^{(i)}_{t + 1} = A^{(i)}x^{(i)}_t + B^{(i)} u^{(i)}_t + w^{(i)}_t
\end{equation}
with the state vector $x^{(i)} \in \mathbb{R}^p$, control input $u^{(i)} \in \mathbb{R}^q$ and the random disturbance a Gaussian noise $w^{(i)}_t \in \mathbb{R}^p$ with covariance matrix $W^{(i)} \in \mathbb{R}^p$. We assume that the pairs $(A^{(i)},B^{(i)})$ are controllable but $A^{(i)}$ is not necessarily stable. 

In wireless control systems, the access point manages the access of each plant to the shared wireless medium. This medium is inherently noisy and prone to packet drops. When the plant is able to successfully receive the control signal, the feedback loop is closed, and the plant can execute the ensuing control action. 
When the plant cannot reliably receive the signal, however, it does not execute any control action.  
The dynamics of each plant can then be represented by
 \begin{equation}
 x^{(i)}_{t + 1} = 
 \begin{cases}
    A^{(i)}x^{(i)}_t + B^{(i)} u^{(i)}_t + w_t, \text{ closed loop},  \\
    A^{(i)}x^{(i)}_t + w_t, \text{ open loop}.
 \end{cases}
 \label{eq:plant_switched}
\end{equation}

\tikzstyle{block} = [draw, fill=blue!20, rectangle, 
    minimum height=1cm, minimum width=5em]
\tikzstyle{largerblock} = [draw, fill=blue!20, rectangle, 
    minimum height=1cm, minimum width=18em]
\tikzstyle{sum} = [draw, fill=blue!20, circle, node distance=1cm]
\tikzstyle{input} = [coordinate]
\tikzstyle{output} = [coordinate]
\tikzstyle{pinstyle} = [pin edge={to-,thin,black}]

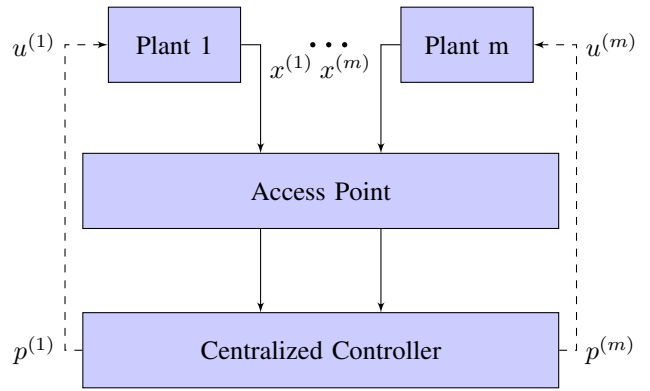
\begin{figure}
 \centering
\begin{tikzpicture}[auto, node distance=2cm,>=latex']
    \node [largerblock] (measurements) {Access Point};
        \node [block, above left of=measurements,
            node distance=2.75cm] (system) {Plant 1};
            \node [block, above right of=measurements,
            node distance=2.75cm] (systemm) {Plant m};
    \node [largerblock, below of=measurements,node distance = 2.12cm](central_controller){Centralized Controller};
    \node [output, above of=measurements, node distance = 0.5cm] (aux1) {};
    \node [output, right of=aux1, node distance = 0.8cm] (teste) {};
     \node [input, right of=teste, node distance = 2.7cm] (teste2) {};
     \node [input, left of=aux1, node distance = 0.8cm] (auxsys1) {};
     \node [input, left of=auxsys1, node distance = 2.7cm] (auxsys1-1) {};
    \draw [->] (system.east) -| node [name=x, below right] {$x^{(1)}$}(auxsys1);
    \draw [->] (systemm.west) -| node [name=xm, below left] {$x^{(m)}$}(teste);
    \node [input, left of=measurements, node distance = 3.4cm] (aux_ap1){};
\node [input, right of=measurements, node distance = 3.4cm] (aux_ap2){};    
    \draw [dashed] [->] (aux_ap1) |- node [name=power1, left] {$u^{(1)}$}(system.west);
    \draw [dashed] [->] (aux_ap2) |- node [name=powerm, right] {$u^{(m)}$}(systemm.east);
    \foreach \i in {0.45,0.55,0.65}
        \fill ($(system.east)!\i!(systemm.west)$) circle[radius=1pt];
    \node [output,below of=measurements, node distance = 0.5cm](aux_below){};
    \node [output, right of=aux_below, node distance = 0.8cm](aux_below_right){};
    \node [output, left of=aux_below, node distance = 0.8cm](aux_below_left){};
    \node [output,above of=central_controller, node distance = 0.5cm](aux_central){};
    \node [output, right of=aux_central, node distance = 0.8cm](aux_central_right){};
    \node [output, left of=aux_central, node distance = 0.8cm](aux_central_left){};
    \draw [->] (aux_below_right) -- (aux_central_right);
    \draw [->] (aux_below_left) -- (aux_central_left);
     \draw [dashed]  (central_controller.west) -| node [name=pow1, left] {$p^{(1)}$} (aux_ap1);
    \draw [dashed] (central_controller.east) -| node [name=powm, right] {$p^{(m)}$ } (aux_ap2);
\end{tikzpicture}
\caption{ Wireless control system architecture. Each plant $i$ has state $x^{(i)}$ and its dynamics is independent of the others. The wireless communication network is made up by different channels  with fading state $h^{(i)}$, and an access manager oversees the communication network. The plants transmit their current states to the access manager, which should then distribute the resources $p^{(i)}$ available in the network to the corresponding communication channels. When the feedback loop is closed, the plant receives the control signal and executes the corresponding control action. When the feedback loop is not closed the plant relies on an estimate of the control signal instead.} 
\label{fig:wl_control_sys} 
\end{figure}

The reliability of the communications channel is dependent upon the resource, or power, level with which a plants sends its information as well as a random channel state known as wireless fading. Let then $h^{(i)} \in \mathcal{H} \subset \mathbb{R}_+$  a random variable drawn from a distribution $m(h)$ that represents the fading state experienced by plant $i$ and denote by $p^{(i)} \in \reals_+$ the resource used by plant $i$. 
Further define a function $q: \reals_{+} \times \reals_+ \rightarrow [0,1]$ that, given a channel state and resource allocation, returns the probability of successfully receiving the packet. The  system dynamics in \eqref{eq:plant_switched} is then given by
\begin{equation}
 x^{(i)}_{t + 1} = 
 \begin{cases}
    A^{(i)}x^{(i)}_t + B^{(i)} u^{(i)}_t + w_t, \text{ w.p. } q(h^{(i)},p^{(i)}),  \\
    A^{(i)}x^{(i)}_t + w_t, \text{ w.p. } 1- q(h^{(i)},p^{(i)}). 
 \end{cases}
 \label{eq:plant_switched_prob}
\end{equation}

As can be seen in the dynamics in \eqref{eq:plant_switched_prob}, allocating more power to a plant will increase the reliability of the wireless channel, thus increasing probability of closing its control loop and experiencing more favorable dynamics.  In most practical systems, however, we do not have unlimited power to allocate between the communication channels. The resource allocation problem consists of properly allocating resources available while keeping all plants in desirable states. 
We are interested in a resource allocation function $p(h,x)$ that, given current channel conditions $h_t := [h_t^{(1)}; \hdots; h_t^{(m)}]$ and plant states $x_t := [x_t^{(1)}; \hdots; x_t^{(m)}]$, distributes resources among the plants without violating a maximum power constraint $p_{max}$. As current resource allocation decisions impact future states, we consider as performance metric a quadratic cost of the plants states evaluated over a finite horizon $T$. 
 Putting all the above pieces together, the constrained resource allocation problem takes the form
\begin{equation}
 \begin{aligned}
  \minimize_{p(h,x)} \, &\mathbb{E}^{p(h,x)}_{x_0}  \left[  \sum_{t = 0}^T x_t^\trp Q_t x_t | x_0 = \hat{x}_0  \right] \\
  \st \, 
  & p(h,x) \in \mathcal{P}; \mathcal{P} = \left\{p(h,x) \, : \, \sum_{i = 1}^m p^{(i)} \leq p_{\max} \right\}, \\
 \end{aligned}
 \label{eq:constrained_optimal_prob}
\end{equation}
with $Q_t \geq 0$ and $p^{(i)}$ the $i$th component of the resource allocation vector $p(h, x)$, i.e. resource allocated to plant $i$. At each time $t$, the AP uses power $p_t^{(i)} = [p(h_t, x_t)]_i$ to send the control signal $u^{(i)}$ back to plant $i$. The communication exchange subsequently occurs with success rate given by $q(h^{(i)}_t, p_t^{(i)})$ and plant $i$ evolves via \eqref{eq:plant_switched_prob} accordingly.

 In \eqref{eq:constrained_optimal_prob}, the objective involves finding the resource allocation function $p(x,h)$ which results in the minimum operation cost of the plants while satisfying the resource constraints. 
Note that this is an infinite-dimensional optimization problem. It is generally intractable to find optimal solutions even for problems with a low number of plants and with short optimization horizons. Moreover, finding an optimal policy directly in \eqref{eq:constrained_optimal_prob} necessarily requires explicit knowledge of the plant dynamics and communication models in \eqref{eq:plant_switched_prob}, which are often unavailable in practice. 
Since we cannot find an optimal solution to the optimal resource allocation problem, we turn to  strategies which can offer feasible but approximate solutions. 
Given recent advances on model-free reinforcement learning, we propose the use of deep
 RL for resource allocation in wireless control systems.  

\section{Reinforcement learning for resource allocation}
\label{sec:resource_RL}
Reinforcement learning can be seen as a mathematical representation of the idea that learning comes from interaction with the environment \cite{sutton_reinforcement_learning,Szepesvri_2010}. 
Reinforcement learning problems are formalized with the use of Markov decision processes (MDP). 
A MDP is a standard mathematical description of a sequential decision making process \cite{sutton_reinforcement_learning}; formally, it consists in a tuple $\left< \mathcal{S},\mathcal{A},\mathcal{P} \right>$ with  $\mathcal{S}$ a  set of states,  $\mathcal{A}$ a set of actions and $\mathcal{P}$ a state transition probability kernel.   The state transition probability kernel  $\mathcal{P} : \mathcal{S} \times \mathcal{A} \times \mathcal{S} \rightarrow [0,1]$ assigns to each triplet $(s,a,s') $ the probability of moving from state $s$ to $s'$ if action $a$ is chosen. 
A transition from a state $s_t$ to $s_{t + 1}$ incurs a cost per stage $r_t$, and the agent takes actions according to a stochastic policy $\pi(a|s)$. That policy corresponds to a distribution that gives the probability of the agent to choose an action $a$ when its current state is $s$.  
The agent's objective is to minimize some cumulative cost 
\begin{equation}
J(\pi, s) :=\mathbb{E}_{s}^\pi \left[ R_t \right] = \mathbb{E}_{s}^\pi \left[ \sum_{k =1}^T \gamma^t r_{k+ t + 1} \right]
\label{eq:rl_reward}
\end{equation} 
starting from a state $s$ and following a policy $\pi(\cdot)$
with $\gamma \in (0,1]$ a given discount factor \cite{hernandez-lerma_discrete-time_1996,sutton_reinforcement_learning,Szepesvri_2010}.
In this setting we define the value function as
\begin{equation}
J^*(s) = \inf_{\pi \in \Pi} J(\pi, s), s \in \mathcal{S},
\label{eq:value_function_def}
\end{equation}
and the problem consists in finding the policy $\pi^*$ that achieves this minimum. 

To cast the scheduling problem of section \ref{sec:prob_formulation} as a reinforcement learning problem, we take the the AP as a centralized agent. 
The actions here correspond to resource allocation vectors, defined on the corresponding action space $\mathbb{R}_+^m$ and taken according to the resource allocation policy $p(h,x)$. 
The possible states of the system are made up by variables describing the state of the plants, $x$, and the channel states $h$, that is
\begin{equation*}
s_t = \left[ h_t;x_t  \right] = \left[ h_t^{(1)}; \hdots; h_t^{(m)}; x_t^{(1)}. \hdots; x_t^{(m)} \right]
\end{equation*}
indicating that the state space in this case is $\mathbb{R}^{m + m}$.
The performance of the resource allocation function is a quadratic cost in the control states, hence we take \eqref{eq:rl_reward} as 
\begin{equation}\label{eqn:value_function_2}
J(\pi, s)  = \mathbb{E}^{p(h,x)}_{s}  \left[  \sum_{t = 0}^{T} \gamma^t x_t^\trp Q_t x_t | x_0 = \hat{x}_0  \right].
\end{equation}
To overcome the challenge of the functional optimization problem, let us parameterize the resource allocation function $p(h,x)$ with some stochastic policy $ \pi (p | s; \theta)$ that is fully specified with a parameter vector $\theta \in \reals^r$, i.e.
\begin{equation}
p(h,x) =  \pi (p | s; \theta). 
\end{equation}
Naturally, the above parameterization incurs a loss of optimality, but multilayer neural networks satisfy the so-called universal approximation  theorem, meaning that these functions can arbitrarily approximate a continuous function when sufficiently large \cite{hornik_1989}. 
With this parameterization we can rewrite the  resource allocation problem in \eqref{eq:constrained_optimal_prob} as
\begin{equation}
 \begin{aligned}
  \minimize_{\theta} \, &\mathbb{E}^{\pi(h,x; \theta)}_{x_0}  \left[  \sum_{t = 0}^T \gamma^t x_t^\trp Q_t x_t | x_0 = \hat{x}_0  \right] \\
  & \pi(h,x;\theta) \in \mathcal{P}; \mathcal{P} = \left\{ \pi(h,x;\theta) \, : \, \sum_{i = 1}^m \pi^{(i)} \leq p_{\max} \right\} \\
 \end{aligned}
 \label{eq:constrained_optimal_prob_param}
\end{equation}
Note that the cost function here will depend on the parameters $\theta$, allowing us to take
\begin{equation}
J(\theta) = \mathbb{E}^{\pi(h,x; \theta)}_{x_0}  \left[  \sum_{t = 0}^T \gamma^t x_t^\trp Q_t x_t | x_0 = \hat{x}_0  \right].
\label{eq:cost_function_params}
\end{equation}

The optimization problem described above is an  example of a model-free reinforcement learning problem: the agent does not know the dynamical model of the control plants or the distribution of the communication channel states nor tries to learn it.
Within the model-free RL framework, an immediate distinction is made between \emph{value-based} methods, that is, methods that learn or estimate the action-value function and calculate the corresponding policy based on this approximation; 
and \emph{policy-based} methods, where the algorithm learns a parameterized policy directly. In the latter case we assume that the policy is differentiable with respect to the parameters $\theta$. 
 Policy-based methods advantages over value function approximation in model-free RL include the possibility of incorporating some prior knowledge about the policy with the chosen parameterization. More importantly, parameterizing a policy directly allows us to learn policies in continuous actions spaces as in the resource allocation problem considered here  \cite[ch. 13]{sutton_reinforcement_learning}. 
 
 Since our objective is to minimize some cost function, at each iteration policy gradient methods perform approximate gradient descent in the cost function $J(\theta)$ \cite[ch. 13]{sutton_reinforcement_learning}, 
\begin{equation}
\theta_{t + 1} = \theta_t - \alpha \nabla \hat{J(\theta_t)},
\end{equation}
where  $\alpha$ is the learning rate and $ \nabla \hat{J(\theta_t)}$ an estimate of the gradient of $J(\theta)$ with respect to $\theta$. Methods which use some variation of this basic gradient descent step are known as policy gradient algorithms. 
The policy gradient theorem gives a closed-form expression for the gradient  $\nabla J(\theta)$ of the cost function $J(\theta)$ with respect to the parameter vector $\theta$ \cite{sutton_reinforcement_learning}. Policy gradient algorithms then come up with strategies to sample the actions, states and rewards of the underlying Markov Decision Process to approximate that policy gradient expression \cite{sutton_reinforcement_learning, Szepesvri_2010}. First we consider the REINFORCE algorithm, where the estimate will depend on the action $a_t$ taken at time $t$ \cite{Williams1992, Sutton2000policygradient},
\begin{equation}
\theta_{t+1} = \theta_t - \alpha \gamma^t R_t \frac{\nabla \pi(a_t|s_t; \theta_t) }{ \pi(a_t|s_t; \theta_t) }.
\label{eq:reinforce_update}
\end{equation}
The equation shows that each update of the REINFORCE algorithm depends on the return $R_t$ associated to the action $a_t$ taken and the ratio between the gradient of the probability of executing that action and the probability of doing so \cite{sutton_reinforcement_learning}. Other policy gradient algorithms follow the same basic structure, although with different estimates of the gradient of the cost function. 
Note that the parameter update on the right-hand side of \eqref{eq:reinforce_update} takes into account only the states, actions and returns sampled from the environment;  
the algorithm does not need to know or to learn a model of the system. 

For resource allocation problems, it is essential that we design policies that not only minimize the control cost in \eqref{eq:cost_function_params}, but do so while satisfying the wireless resource constraints defined by $\mathcal{P}$ in \eqref{eq:constrained_optimal_prob_param}.  In particular, we may select or design the stochastic policy distribution $\pi( p | s; \theta)$ to output resource allocation actions $p$ such that $\sum p_i \leq p_{\text{max}}$ by construction. This amounts to outputing actions that are normalized, or belong to a $m-1$ simplex, and scaling by $p_{\max}$. Natural choices for such a distribution include a Dirichlet distribution parameterized by $\theta$ or a series of independent Gaussian distributions parameterized by $\theta_i$. The latter case would require some normalization procedure, such as softmax, to properly scale the power allocations.  

\begin{figure}
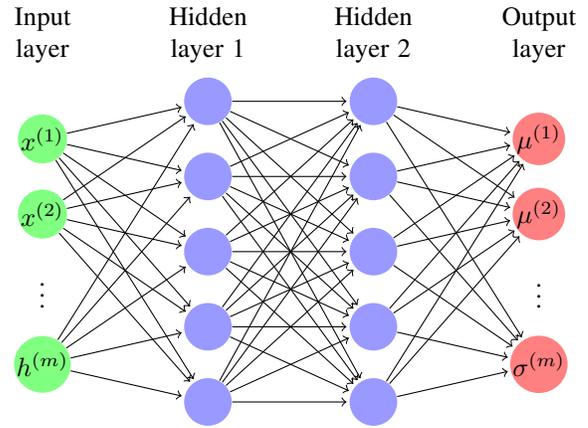
 \centering
\begin{neuralnetwork}[nodespacing=10mm, layerspacing=22mm,
			maintitleheight=2.5em, layertitleheight=2.5em,
			height=5, toprow=false, nodesize=18pt, style={},
			title={}, titlestyle={}]
    \newcommand{\nodetextclear}[2]{}
    \newcommand{\nodetextx}[2]{\ifnum #2=4 $h^{(m)}$ \else $x^{(#2)}$ \fi}
    \newcommand{\nodetexty}[2]{\ifnum #2=4 $\sigma^{(m)}$ \else $\mu^{(#2)}$ \fi}
    \inputlayer[count=4, bias=false, exclude={3}, title=Input\\layer, text=\nodetextx]
    \hiddenlayer[count=5, bias=false, title=Hidden\\layer 1, text=\nodetextclear]
      \linklayers[not from={3}]
      \hiddenlayer[count=5, bias=false, title=Hidden\\layer 2, text=\nodetextclear]
      \linklayers
    \outputlayer[count=4, exclude={3}, title=Output\\layer, text=\nodetexty] 
      \linklayers[not to={3}]

    \path (L0-2) -- node{$\vdots$} (L0-4);
    \path (L3-2) -- node{$\vdots$} (L3-4);
\end{neuralnetwork}
 \caption{Basic architecture of a neural network. Each node in the first hidden layer takes a linear combination of the nodes in the input layer and produces a nonlinear transformation of this linear combination. The nonlinear transformation is given by an activation function, usually a sigmoid or rectified linear function. The nodes in the second hidden layer take a linear combination of the outputs of the first hidden layer and once again produce a nonlinear transformation of that linear combination. The process is repeated until the output layer.  Here the inputs correspond to the plant states and channel variables, and the output is a set of parameters used to characterize the allocation policy.}
\label{fig:basic_nn}
\end{figure}

The choice of parameters for the policy in a reinforcement learning problem is flexible, and that parameterization allows us to search for optimal policies within a certain class of functions.  
 Resource allocation functions such as the one we want to approximate here, however, do not necessarily have a known form. Thus we leverage neural networks to search for allocation policies within a larger class of nonlinear functions; cf. Figure \ref{fig:basic_nn} for a representation of neural networks. 
 For the resource allocation problem studied here, the neural network takes the plants states $x^{(1)}_t, \dots, x^{(m)}_t$ and channel variables $h^{(1)}_t, \dots, h^{(m)}_t$ 
  as inputs, and outputs a set of parameters used to characterize a (multivariate) Gaussian policy with means $\mu^{(1)}, \dots,\mu^{(m)} $ and covariance matrices $\sigma^{(1)}, \dots,\sigma^{(m)} $.
In a multilayer, fully connected neural network, each element in the first hidden layer constructs a linear combination of the input elements and passes this combination through a nonlinear transformation or activation function $\phi(\cdot)$. 
Each element in the second hidden layer then performs a similar transformation on the elements of the first hidden layer, and the process is repeated up to the output layer. 
At each hidden layer $l$ this 
generates the so-called hidden units \cite[ch. 5]{Bishop_2006}
 \begin{equation}
z_{l} = \phi( C_{l} z_{l-1} + b_l). 
 \end{equation}
 Here the matrix $C_l$ and the vector $b_l$  
 represent the linear combination. 
The outputs $y^{(k)}$ of the neural network will then be given by \cite{Bishop_2006}
\begin{equation}
y^{(k)}(x,h) = \phi \left(  \phi \left( \dots \phi \left( C_1 z_0 + b_1 \right) \right) + b_L  \right) 
\label{eq:nn_output}
\end{equation}
with $z_0 = [x;h]$ and $y^{(L)}(x,h) = [\mu; \sigma]$, while $d_l$ is the number of hidden units in hidden layer $l = 1, \dots, L$ and the parameter vector is given by $\theta = [C_1; b_1; \hdots; C_L; b_l]$.
 The activation function must be differentiable and nonlinear (otherwise we would retrieve a standard linear combination of the inputs in the neural network), and common choices include rectified linear units (ReLU), hyperbolic tangent and sigmoid functions. We consider here ReLU activation functions, since they handle better issues such as vanishing gradients \cite{glorot_deep_ReLU}.
 Policy-based deep RL algorithms use a neural network to represent the parameters of the policy distribution.

\begin{algorithm}[h]
     \caption{Deep PG for resource allocation in wireless control systems (based on  \cite{sutton_reinforcement_learning})}
     \label{alg:deep_pg}
   \label{alg:dnn_wcs}
  \DontPrintSemicolon
  \SetKwInOut{Required}{Required}
  \Required{System dynamics (to generate episodes); cost objective $J(\cdot)$; horizon $T$; number of episodes $N$.}
   \KwResult{Offline resource allocation algorithm.}
      \BlankLine
   initialization: load initial training / parameter set $\Theta$ \;
   \tcc{Loop over the episodes}
   \For{$ii$ = $1, \dots, N$}{
\tcc{$T$-step horizon simulation}
generates complete episode: \;
$x_0$, $h_0$, $p_0, r_1, \dots, x_{T-1}, h_{T-1}, p_{T-1}, r_{T}$\;
   \While{ t < T}{
   	\tcc{calculates cost-to-go}
   	$R_t$ $\leftarrow \sum_{k = t + 1}^T \gamma^{k - t - 1} r_k $ \; 
   	\tcc{updates parameters}
   	$\theta \leftarrow \theta - \alpha \gamma^t R_t \nabla \ln \pi(p_t|s_t; \theta)  $ \;
	}
   }
  \end{algorithm}


\section{Numerical experiments}
\label{sec:num_exp}
We now present numerical experiments to illustrate the use of the proposed learning-based approach. 
First we consider the distribution of a power budget among a collection of unstable but controllable scalar plants sharing a wireless communication network when state information is available without noise.
 The probability of successfully receiving an information packet is given by a function $q \, : \, \mathbb{R}_+ \times \mathbb{R} \to [0,1] $ depending on the channel estimate and allocated power. 
Here we considered an exponential distribution
\begin{equation}
q^{(i)}(h, p) = 1 - \exp \left(- h^{(i)} p^{(i)}(h,x) \right).
\end{equation}
This distribution gives us the probability of the corresponding feedback loop closing at a given time instant.
 For the numerical experiments we consider that the controller does not act when the transmission fails. 
For the policy-based case, we use a standard REINFORCE approach \cite{Williams1992, Sutton2000policygradient}. 

\begin{figure}
\includegraphics[width=\linewidth]{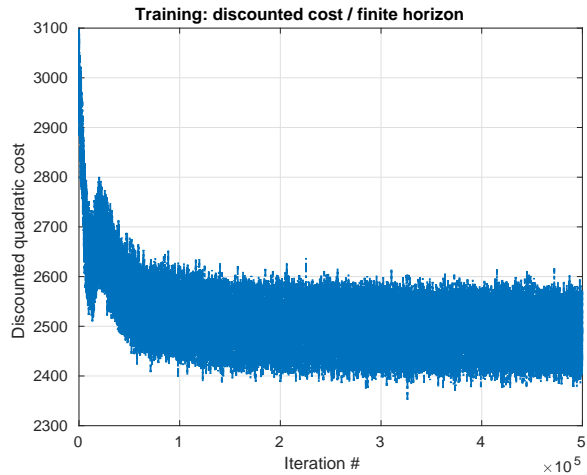}
\caption{Discounted cost (for a full episode) during learning phase. This simulation considers 15 unstable but controllable scalar plants. }
\label{fig:training_cost}
\end{figure}

\begin{figure}
\includegraphics[width=0.5\textwidth]{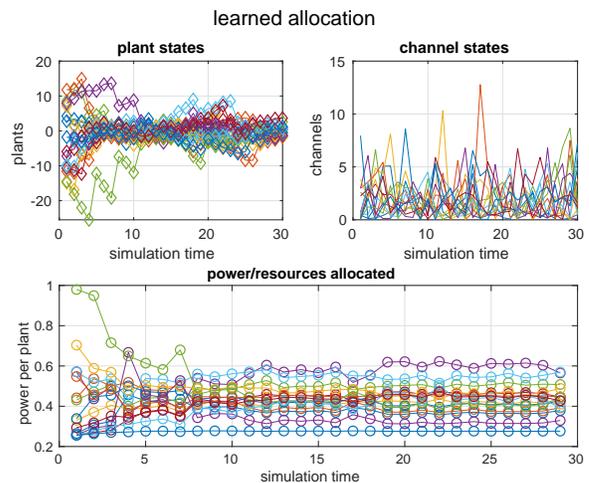}
\caption{Learned resource allocation policy: plant states, channel conditions and allocated power.}
\label{fig:learned_allocation_states}
\end{figure}

The numerical results presented here consider 15 unstable, randomly chosen scalar plants sharing a total power budget $p_{\text{max}} = 6$. Training was taken with 1000 samples per iteration and considering an optimization horizon of $T = 5$, whereas the test and comparison with some naïve benchmarks (dividing power equally among all the control plants or giving more power to the plant further away from the equilibrium point) were performed with $T = 30$. Figure \ref{fig:training_cost} shows the overall discounted cost for training episodes over iterations of the learning procedure; as expected, the performance of the learned policy improves as more experience is collected. Nonetheless, the plot also shows that the learning process is fairly noisy. This is in part due to the fact that REINFORCE has high variance and needs a large number of samples to achieve good results \cite{sutton_reinforcement_learning}.

\begin{figure}
\includegraphics[width=\linewidth]{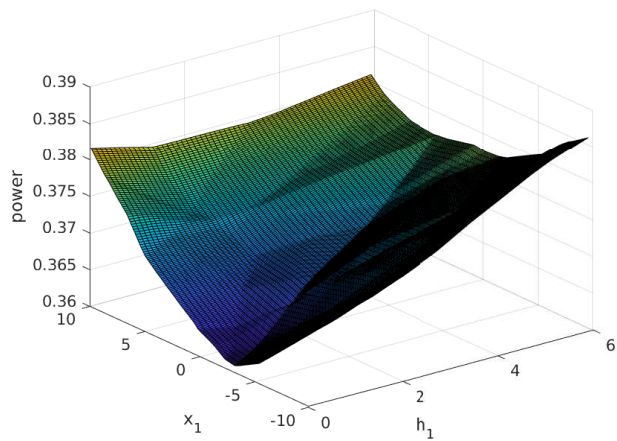}
\caption{Resource allocation with 15 scalar plants: learned policy for the first plant.
}
\label{fig:learned_allocation_policy}
\end{figure}

Figure \ref{fig:learned_allocation_states} illustrates the learned allocation policy for the same simulation. The plot shows the evolution of the plants and channels states, and the corresponding power allocation. 
 As expected, the most unstable plants at a given time instant receive more power. 
   That behavior  can be seen more clearly in Figure \ref{fig:learned_allocation_policy}, which shows how the learned policy allocates power to the first plant for different state and channel conditions while the other plants are kept at $x^{(i)} = 0$.

\begin{figure}
\includegraphics[width=\linewidth]{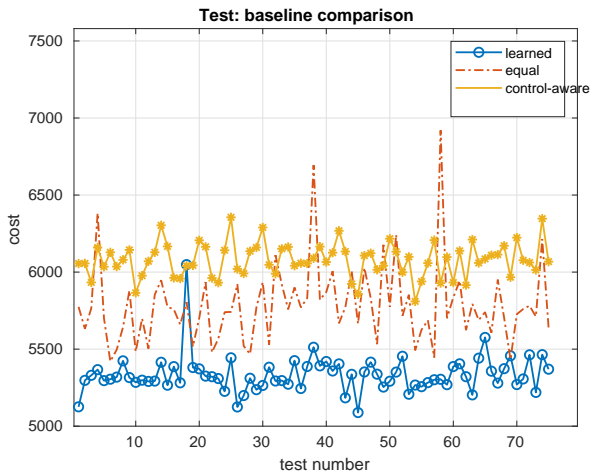}
\caption{Resource allocation with 15 scalar plants: test compariso between the learned policy (blue), equal power allocation between the plants (orange) and a simple control-aware heuristic (yellow). 
}
\label{fig:test_comp}
\end{figure}

After the training phase, the learned allocation policy was compared to the benchmarks mentioned above. The test was executed with a larger horizon than the training phase, in order to see how the learned policy would adapt to a longer implementation setting.  The plant dynamics and control policies were kept the same as in the training phase. Figure \ref{fig:test_comp} shows the results obtained. 
 The learned allocation policy (in blue) was able to get better results than the benchmarks it was compared against in almost all the simulation scenarios. 

\begin{figure}
\includegraphics[width=\linewidth]{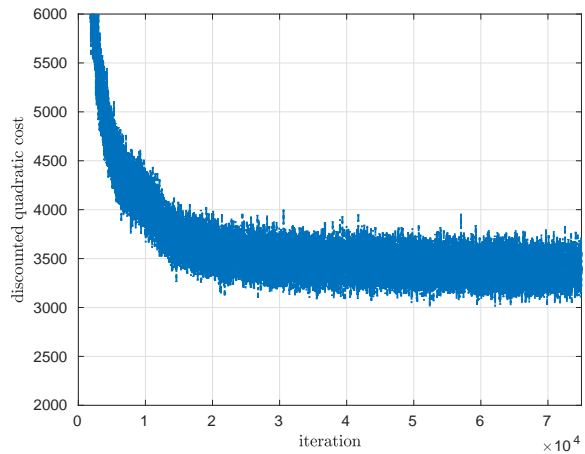}
\caption{Discounted cost (for a full episode) during learning phase. This simulation considers 10 unstable but controllable scalar plants. Here the agent has access only to noisy observations of the plants and channel states. }
\label{fig:noisy_training_cost}
\end{figure}

The second test considered a more challenging scenario where 10 scalar plants share a power budget of $p_{\text{max}} = 3$, resulting in a smaller average power per plant. The channels fading states follow an exponential distribution with parameter $\lambda_h = 1$. Here we consider a more realistic setting in which the AP has access only to noisy estimates of the control states and channel conditions, i.e. the allocation decision is taken based on
\begin{equation}
[ \tilde{h_t}; \tilde{x_t} ] =  \left[ h_t;x_t  \right] + w_t^{(o)}
\end{equation}
where the observation noise $w_t^{(o)}$ is a Gaussian disturbance with covariance $W^{(\text{obs})}$. Training was performed with a horizon $T = 10$ and test with $T = 40$. We used 300 samples per iteration. The neural network was initialized with a supervised pre-learning phase to initially fit a heuristic that gives more power to more unstable plants. Figure \ref{fig:noisy_training_cost} shows the training cost for this simulation, where we considered $W^{(\text{obs})} = 0.4$.  

\begin{figure}
\includegraphics[width=\linewidth]{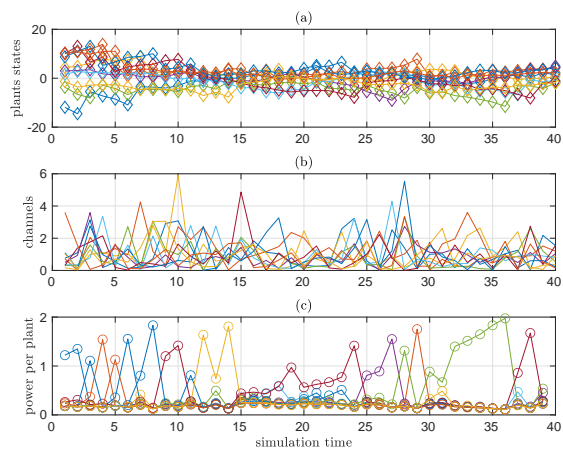}
\caption{Test phase example for the learned allocation. Figure shows (a) the evolution of the plant states, (b) the channel fading conditions and (c) the power allocated to each plant. }
\label{fig:noisy_test_example}
\end{figure}

\begin{figure}
\includegraphics[width=\linewidth]{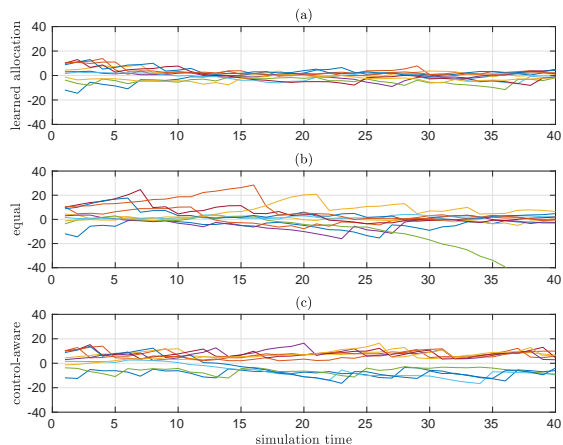}
\caption{Test phase example. Figure shows  the evolution of the plant states for (a) the learned allocation, (b) dividing power equally among the plants and (c) the control-aware heuristic. }
\label{fig:noisy_test_example_states}
\end{figure}

\begin{figure}
\includegraphics[width=\linewidth]{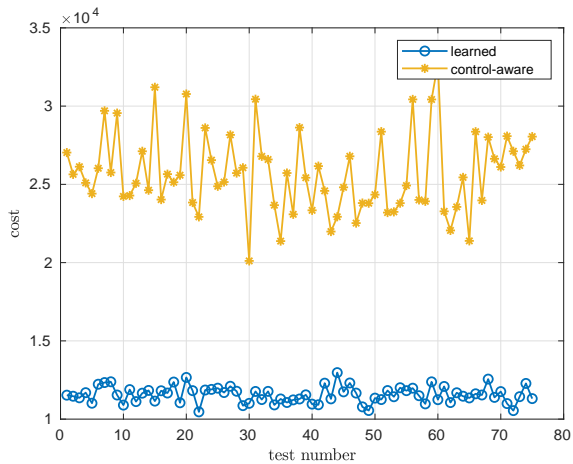}
\caption{Test comparison between the learned allocation (blue) and a control-aware heuristic (yellow) (dividing power equally among the plants performed significantly worse in this case).}
\label{fig:noisy_test_comparison}
\end{figure}

Figure \ref{fig:noisy_test_example} shows an example of the test scenario after the policy was learned; as expected, we see that the learned policy keeps giving more power to more unstable plants. Figures \ref{fig:noisy_test_example_states} and  \ref{fig:noisy_test_comparison} bring a comparison between the learned policy and the heuristics mentioned before. 
Note that in this setting the improvement of the learned policy upon the best baseline is larger than in the previous setting: here the total cost of the learned policy hovers around 12000, whereas that value hovers around 25000 for the control-aware heuristic. 
It is important to point out that REINFORCE is a simple reinforcement learning algorithm, and we expect to get better results with higher sample efficiency when implementing more recent, state-of-the-art actor-critic algorithms. 

\section{Conclusion}

This paper discusses a deep reinforcement learning approach for resource allocation in wireless control systems. On the one hand, resource allocation problems are usually hard to solve, so it is natural to leverage heuristics to find an approximate allocation policy. Deep reinforcement learning algorithms, on the other, have achieved good results in traditional AI benchmarks, which, combined with their model-free learning capabilities, make it an attractive framework to handle resource allocation problems in wireless control systems where explicit system information is often unknown. 

Here we made use of policy-based deep RL algorithms that allow us to learn continuous allocation policies based on current control and channel state information.  
Numerical results presented here show that the proposed approach outperforms baseline resource allocation policies. In future work, we plan 
to make use of more sophisticated deep reinforcement learning algorithms to improve performance and sample efficiency.


\bibliographystyle{unsrt}
\bibliography{wl_control}

\end{document}